\documentclass{npcs}
\usepackage{graphicx,rotating}
\usepackage{amsmath, amssymb}
\usepackage{axodraw}
\def\TeV{\ifmmode {\,\mathrm{ Te\kern -0.1em V}}\else
                   \textrm{Te\kern -0.1em V}\fi}%
\def\GeV{\ifmmode {\,\mathrm{ Ge\kern -0.1em V}}\else
                   \textrm{Ge\kern -0.1em V}\fi}%
\def\MeV{\ifmmode {\,\mathrm{ Me\kern -0.1em V}}\else
                   \textrm{Me\kern -0.1em V}\fi}%
\def\keV{\ifmmode {\,\mathrm{ ke\kern -0.1em V}}\else
                   \textrm{ke\kern -0.1em V}\fi}%
\def\eV{\ifmmode  {\,\mathrm{ e\kern -0.1em V}}\else
                   \textrm{e\kern -0.1em V}\fi}%

\newcommand{\pb}{\,\rm{pb}}

\newcommand{\lunit}{\mathrm{cm}^{-2}\mathrm{s}^{-1}}
\newcommand{\MH}      {m_{\mathrm{H}}}
\newcommand{\bb}    {{\mathrm b\bar{\mathrm b}}}
\newcommand{\ee}    {\mathrm{e}^+\mathrm{e}^-}

\begin{document}

\runauthor{V. Makarenko, K. M\"onig, T. Shishkina}
\runtitle{
Measuring the Luminosity of a $\gamma \gamma$ Collider with 
$\gamma \gamma \rightarrow \ell^+ \ell^- \gamma$ Events
}

\begin{topmatter}
\begin{flushright}
LC-PHSM-2003-016\\
June 16, 2003
\end{flushright}
%
\boldmath
\title{
Measuring the Luminosity of a $\gamma \gamma$ Collider with 
$\gamma \gamma \rightarrow \ell^+ \ell^- \gamma$ Events
}
\unboldmath
\author{V. Makarenko}
\institution{NC PHEP BSU}
\email{makarenko@hep.by}
\author{K. M\"onig}
\institution{DESY, Zeuthen}
\email{klaus.moenig@desy.de}
\author{T. Shishkina}
\institution{NC PHEP BSU}
\email{shishkina@hep.by}
\vspace{1cm}
\begin{abstract}
The process $\gamma \gamma \rightarrow \ell^+ \ell^-$ is highly
suppressed when the total angular momentum of the two colliding photons is 
zero so that it cannot be used for luminosity determination. This
configuration, however is needed for Higgs production at a photon
collider. It will be shown that the process $\gamma \gamma \rightarrow
\ell^+ \ell^- \gamma$ can be used in this case to measure the
luminosity of a collider with a precision that is good enough not to
limit the error on the partial decay width 
$\Gamma( {\rm H} \rightarrow \gamma \gamma)$.
\end{abstract}  
\end{topmatter}
\mathchardef\vm="117
\mathchardef\um="11D
\mathchardef\E="245
\mathchardef\Mom="250
\def\z1{z{}'}
\def\t1{t{}'}
\def\u1{u{}'}
\def\s1{s{}'}
\def\m2{m^2}
\def\d12{\frac{1}{2}}
\def\lfr#1#2{\ln{\frac{#1}{#2}}}

\def\ub{\bar{\um}}
\def\Jint#1{\mathcal{J} ( {#1} )}
\def\Iint#1#2{\int\limits_{0}^{\ub}{\Jint{#1}} {#2} {d \um}}
\def\Iintl#1#2{\int\limits_{2 m \lambda}^{\ub}{\Jint{#1}} {#2} {d \um}}

\def\bracket#1{\left({#1}\right)}
\def\bra#1{\bracket{#1}}
\def\spr#1#2{\, #1 \! \cdot \! #2 \,}
\def\ddx#1{\frac{1}{#1}}

\newpage

\section{Introduction}

Linear lepton colliders will provide
the possibility to investigate
photon collisions at energies and luminosities close to
those in $e^+e^-$ collisions \cite{gg_proposal}.
If a light Higgs exists one of the main tasks of a photon collider
will be the measurement of the partial width 
$\Gamma( {\rm H} \rightarrow \gamma \gamma)$ \cite{ggtdr}.
Not to be limited by the error from the luminosity determination the
luminosity of the collider at the energy of the Higgs mass has to be
known with a precision of around 1\%. To produce scalar Higgses the
total angular momentum of the two photons has to be $J=0$. In this
case the cross section $\gamma \gamma \rightarrow \ell^+ \ell^-$
is suppressed by a factor $m_\ell^2/s$ and thus not usable for
luminosity determination.
The radiative process
$\gamma \gamma \rightarrow \ell^+ \ell^-\gamma$ is suppressed by an
additional factor $\alpha$, however due to the additional final state
photon the spin suppression does no longer apply. It is therefore
worth examining if the radiative process can be used for luminosity
determination. 

For this reason we consider the exclusive reaction $\gamma\gamma\to
\ell^+ \ell^- \gamma$ as a possible candidate 
for a calibration channel at a photon collider.

The two helicity
configurations of the $\gamma \gamma$-system lead to different spectra
of the final state particles.
We have analysed the behaviour of the
$\gamma\gamma\to \ell^+ \ell^-\gamma$ reaction
for the various helicities of the beam
as a function of different observables.
Since the photon beams in the collider are only partially polarised
the ratio of the cross sections of $\gamma \gamma \to \ell^+ \ell^- \gamma$ 
scattering for $J=$ to $J=2$-beams should be high
for the luminosity measurement.
The main emphasis of this analysis is put on $\sqrt{s}=120 \GeV$, which is 
about the mass, where a light Higgs boson is expected.

\boldmath
\section{Cross sections of $ \gamma\gamma\to \ell^+ \ell^- \gamma$}
\unboldmath

We consider the process
\begin{eqnarray}
\gamma(p_1, \lambda_1)+\gamma(p_2, \lambda_2) \to f(p_1{}', e_1{}') + \bar{f}(p_2{}', e_2{}') + \gamma(p_3, \lambda_3),
\end{eqnarray}
where $\lambda_{i}$ and $e_{i}{}'$ are photon and fermion helicities.

The centre of mass energy squared is denoted by 
$s = {\left(p_1+p_2\right)}^2 = 2 \spr{p_1}{p_2},$
and the final-state photon energy by $\omega=p_3$.
For the differential cross-section we introduce the normalised
final-state photon energy (c.m.s. is used) 
$x=w\slash\sqrt{s}$.
The total helicity of the $\gamma \gamma$ system is denoted by 
$J=|\lambda_1 - \lambda_2|$.

We consider the cross section
\begin{eqnarray}\nonumber
\sigma =\int \ddx{2 s} {\left|M (\lambda_1,\lambda_2,e_1{}',e_2{}',\lambda_3)\right|}^{2} d \Gamma,
\end{eqnarray}
where the phase-space volume element is defined by
\begin{eqnarray}\nonumber
d \Gamma =
\frac{d^{3} p_1'}{\bra{2 \pi}^{3} 2 p_1'{}^0} 
\cdot \frac{d^{3} p_2'}{\bra{2 \pi}^{3} 2 p_2'{}^0} 
\cdot \frac{d^{3} p_3}{\bra{2 \pi}^{3} 2 p_3^0}
\cdot {\bra{2 \pi}^{4}} \delta \bra{p_1+p_2-p_1'-p_2'-p_3}.
\end{eqnarray}

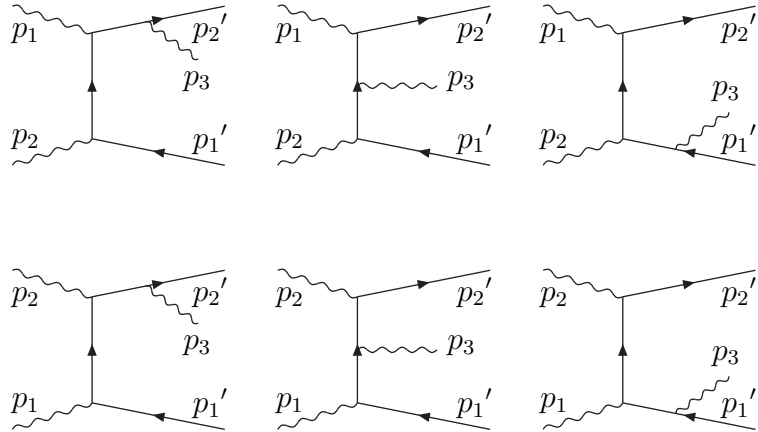
\begin{figure}[htb]
\begin{picture}(300,200)(-90,0)
\Photon(10, 20)(40, 30){1}{4}
\Photon(10, 80)(40, 70){1}{4}
\ArrowLine(90, 20)(40, 30)
\ArrowLine(40, 30)(40, 70)
\ArrowLine(40, 70)(90, 80)
\Photon(60, 74)(80, 60){1}{4}
\Text(15,30)[c]{\normalsize $p_1$}
\Text(15,70)[c]{\normalsize $p_2$}
\Text(80,52)[c]{\normalsize $p_3$}
\Text(85,30)[c]{\normalsize $p_1{}'$}
\Text(85,72)[c]{\normalsize $p_2{}'$}

\Photon(110, 20)(140, 30){1}{4}
\Photon(110, 80)(140, 70){1}{4}
\ArrowLine(190, 20)(140, 30)
\ArrowLine(140, 30)(140, 70)
\ArrowLine(140, 70)(190, 80)
\Photon(140, 50)(170, 50){1}{4}
\Text(115,30)[c]{\normalsize $p_1$}
\Text(115,70)[c]{\normalsize $p_2$}
\Text(180,52)[c]{\normalsize $p_3$}
\Text(185,30)[c]{\normalsize $p_1{}'$}
\Text(185,72)[c]{\normalsize $p_2{}'$}

\Photon(210, 20)(240, 30){1}{4}
\Photon(210, 80)(240, 70){1}{4}
\ArrowLine(290, 20)(240, 30)
\ArrowLine(240, 30)(240, 70)
\ArrowLine(240, 70)(290, 80)
\Photon(260, 26)(280, 40){1}{4}
\Text(215,30)[c]{\normalsize $p_1$}
\Text(215,70)[c]{\normalsize $p_2$}
\Text(280,48)[c]{\normalsize $p_3$}
\Text(285,30)[c]{\normalsize $p_1{}'$}
\Text(285,72)[c]{\normalsize $p_2{}'$}

\SetOffset(0,100)
\Photon(10, 20)(40, 30){1}{4}
\Photon(10, 80)(40, 70){1}{4}
\ArrowLine(90, 20)(40, 30)
\ArrowLine(40, 30)(40, 70)
\ArrowLine(40, 70)(90, 80)
\Photon(60, 74)(80, 60){1}{4}
\Text(15,30)[c]{\normalsize $p_2$}
\Text(15,70)[c]{\normalsize $p_1$}
\Text(80,52)[c]{\normalsize $p_3$}
\Text(85,30)[c]{\normalsize $p_1{}'$}
\Text(85,72)[c]{\normalsize $p_2{}'$}

\Photon(110, 20)(140, 30){1}{4}
\Photon(110, 80)(140, 70){1}{4}
\ArrowLine(190, 20)(140, 30)
\ArrowLine(140, 30)(140, 70)
\ArrowLine(140, 70)(190, 80)
\Photon(140, 50)(170, 50){1}{4}
\Text(115,30)[c]{\normalsize $p_2$}
\Text(115,70)[c]{\normalsize $p_1$}
\Text(180,52)[c]{\normalsize $p_3$}
\Text(185,30)[c]{\normalsize $p_1{}'$}
\Text(185,72)[c]{\normalsize $p_2{}'$}

\Photon(210, 20)(240, 30){1}{4}
\Photon(210, 80)(240, 70){1}{4}
\ArrowLine(290, 20)(240, 30)
\ArrowLine(240, 30)(240, 70)
\ArrowLine(240, 70)(290, 80)
\Photon(260, 26)(280, 40){1}{4}
\Text(215,30)[c]{\normalsize $p_2$}
\Text(215,70)[c]{\normalsize $p_1$}
\Text(280,48)[c]{\normalsize $p_3$}
\Text(285,30)[c]{\normalsize $p_1{}'$}
\Text(285,72)[c]{\normalsize $p_2{}'$}
\end{picture}
\caption{
Diagrams for the process $ \gamma\gamma\to \ell^+ \ell^-\gamma$.
}
\label{fig:ffgfeyn}
\end{figure}

The process $ \gamma\gamma\to \ell^+ \ell^-\gamma$ is
in Born approximation a pure QED reaction. The contributing Feynman
diagrams are shown in figure \ref{fig:ffgfeyn}.
Using the method of helicity amplitudes \cite{ha}, the squared matrix 
elements are obtained \cite{gg_2fg}:
\begin{eqnarray}\label{br3}
{\left|M^{+--++}\right|}^{2} = 4 e^6
\frac{\spr{p_1{}'\!}{p_2{}'\!} \bra{\spr{p_2{}'\!}{p_2}}^{2}}{\spr{p_1{}'\!}{p_3} \spr{p_2{}'\!}{p_3} \spr{p_1{}'\!}{p_1} \spr{p_2{}'\!}{p_1}}.
\end{eqnarray}

All other non-vanishing amplitudes are
obtained from $\left|M^{+--++}\right|$
by using C, P, Bose and crossing (between final and initial particles) symmetries:
\begin{eqnarray*}
d \sigma^{+-+--} & = &
d \sigma^{+--++} {}_{\mid_{1\leftrightarrow 2}}, \; ({\rm P+Bose}) \\
d \sigma^{+-+-+} & = &
d \sigma^{+--++} {}_{\mid_{1'\leftrightarrow 2'}}, \;({\rm C})\\
d \sigma^{+--+-} & = &
d \sigma^{+--++} {}_{\mid_{1\leftrightarrow 2 \atop
1'\leftrightarrow 2'}}, \;({\rm CP+Bose})\\
d \sigma^{+++--} & = &
d \sigma^{+--++} {}_{\mid_{3\leftrightarrow 2 \atop
1'\leftrightarrow 2'}}, \;({\rm C+crossing})\\
d \sigma^{++-+-} & = &
d \sigma^{+++--} {}_{\mid_{1'\leftrightarrow 2'}},
\;({\rm C})\\
d \sigma^{-\lambda_1,-\lambda_2,-e_1{}',-e_2{}',-\lambda_3} & = &
d \sigma^{\lambda_1,\lambda_2,e_1{}',e_2{}',\lambda_3}.
\;({\rm P})
\end{eqnarray*}

Since final-state polarisations cannot be measured we sum
over all final particle helicities.
The further integration is performed numerically using a
Monte-Carlo method \cite{mc}.

\boldmath
\section{Comparison of $J\!=\!0$ and $J\!=\!2$ contributions}
\unboldmath

Calculations for various experimental restrictions on
the parameters of final particles have been performed.
The events are not detected if energies and angles are below the 
corresponding threshold values.
The considered cuts on the phase-space of final particles
are denotes as follows:
\begin{itemize}
\item Minimum final-state photon energy: $\omega_{min}$,
\item Minimum fermion energy: $E_{\ell,min}$,
\item  Minimum angle between any final and any initial state particle
  (polar angle cut): $\Theta_{min}$,
\item Minimum angle between any pair of final state particles: $\varphi_{min}$.
\end{itemize}



\begin{figure}[p]
\leavevmode
\centering
\includegraphics[width=0.48\linewidth, bb=5 25 292 285, angle=0]{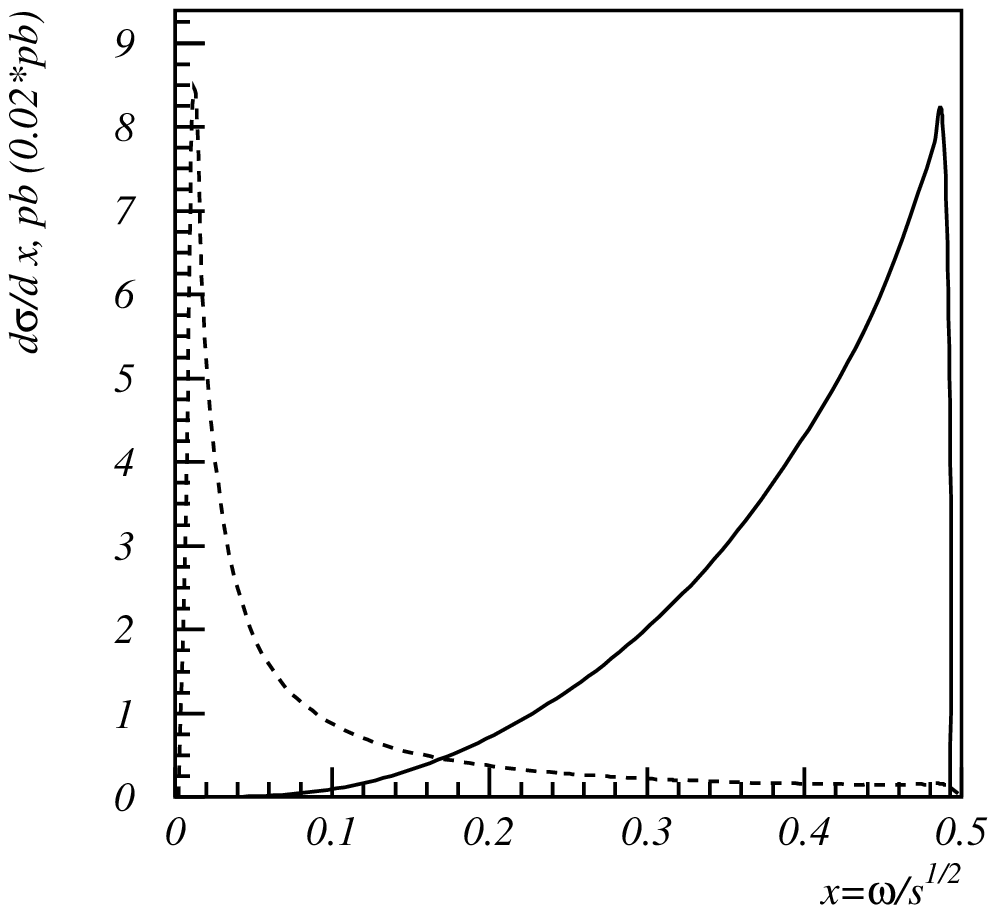} 
\includegraphics[width=0.48\linewidth, bb=5 25 292 285, angle=0]{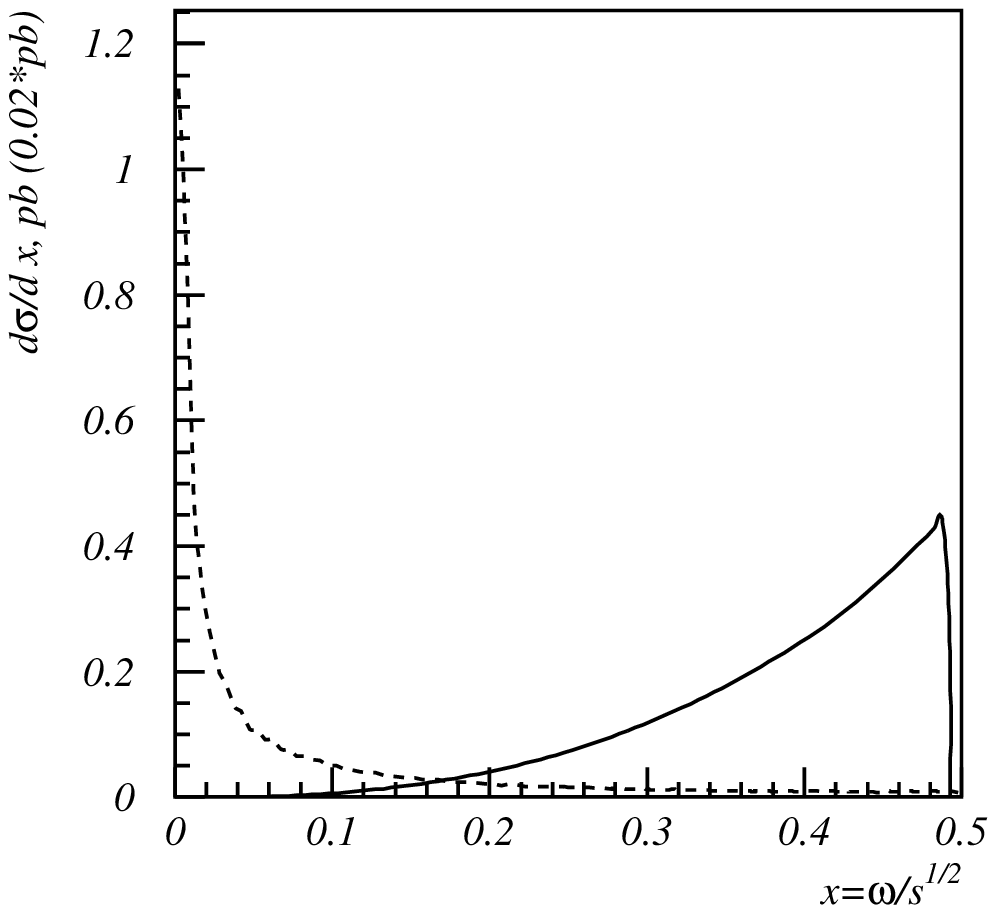}
\caption{
Final-state photon energy spectrum for $J\!=\!0$ (dotted) and 
$J\!=\!2$ (solid)
at $\sqrt{s}=120\GeV$ (left) and $\sqrt{s}=500\GeV$ (right).
Cuts: $\Theta_{min} \!=\! 7^\circ$, $\varphi_{min} \!=\! 3^\circ$, $E_{\ell, min} \!=\! 1 \GeV$,
$\omega_{min} \!=\! 1 \GeV$. 
The $J=2$ cross section has been multiplied by 0.02.
}\label{fig_2}
\end{figure}

\begin{figure}[p]
\leavevmode
\centering
\includegraphics[width=0.48\linewidth, bb=5 25 292 285, angle=0]{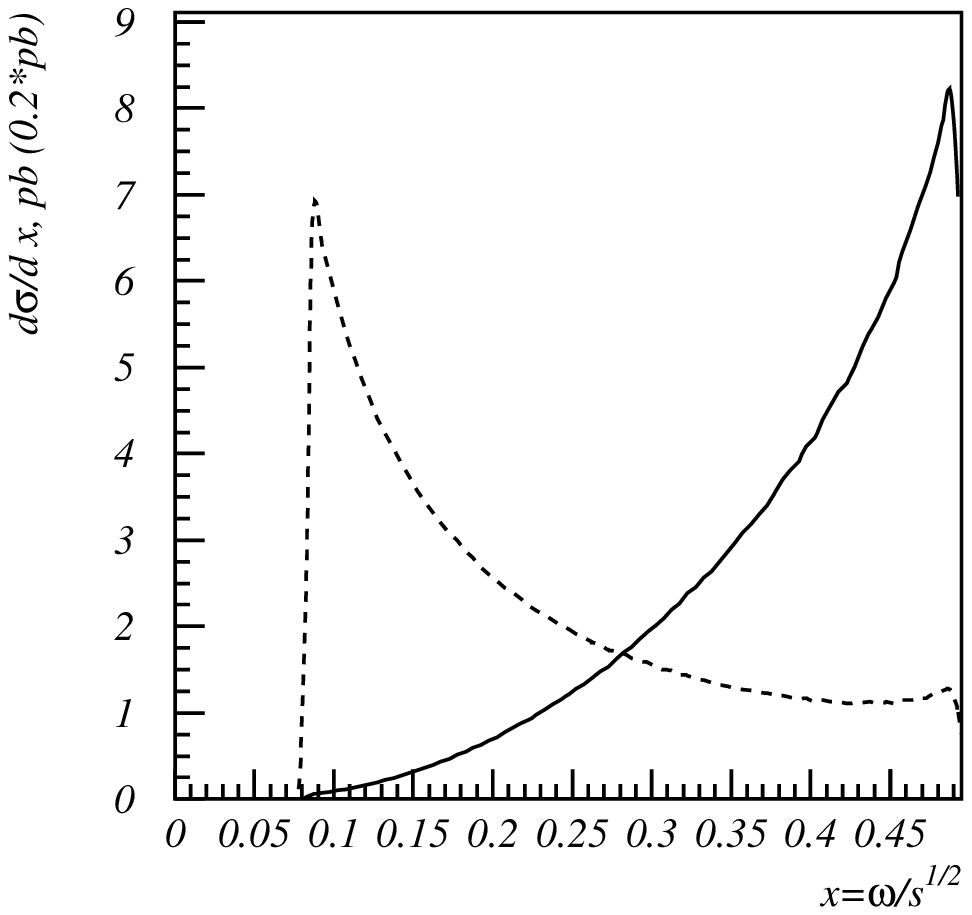}
\includegraphics[width=0.48\linewidth, bb=5 25 292 285, angle=0]{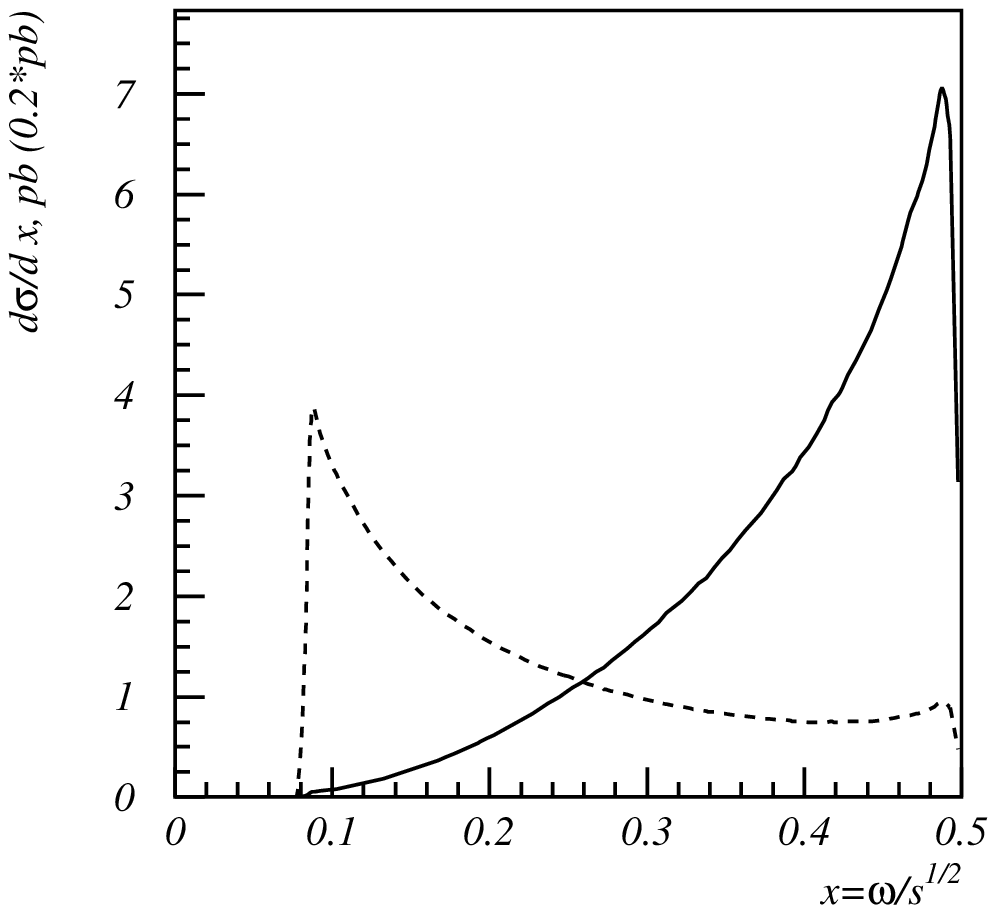}
\caption{
Final-state photon energy spectrum for $J\!=\!0$ (solid) and 
$J\!=\!2)$ (dotted)
at $\sqrt{s}=120\GeV$. Cuts:
$\Theta_{min} \!=\! 7^\circ$, $\varphi_{min} \!=\! 10^\circ$ (left) and $\varphi_{min} \!=\! 30^\circ$ (right),
$E_{\ell, min} \!=\! 1 \GeV$, $\omega_{min} \!=\! 10 \GeV$.
The $J=2$ cross section has been multiplied by 0.2.
}\label{fig_3}
\end{figure}

In figures \ref{fig_2} and \ref{fig_3} the photon energy spectra for 
the beam polarisations $J\!=\!0$ and $J\!=\!2$ at various cuts are presented .
The differential cross section
${d \sigma}\slash{d x}$ for $J=2$ beams
decreases while the one for $J=0$ beams rises with increasing 
final-state photon energy.


\begin{figure}[h!]
\centering
\leavevmode
\includegraphics[width=0.32\linewidth, bb=8 21 234 342, angle=0]{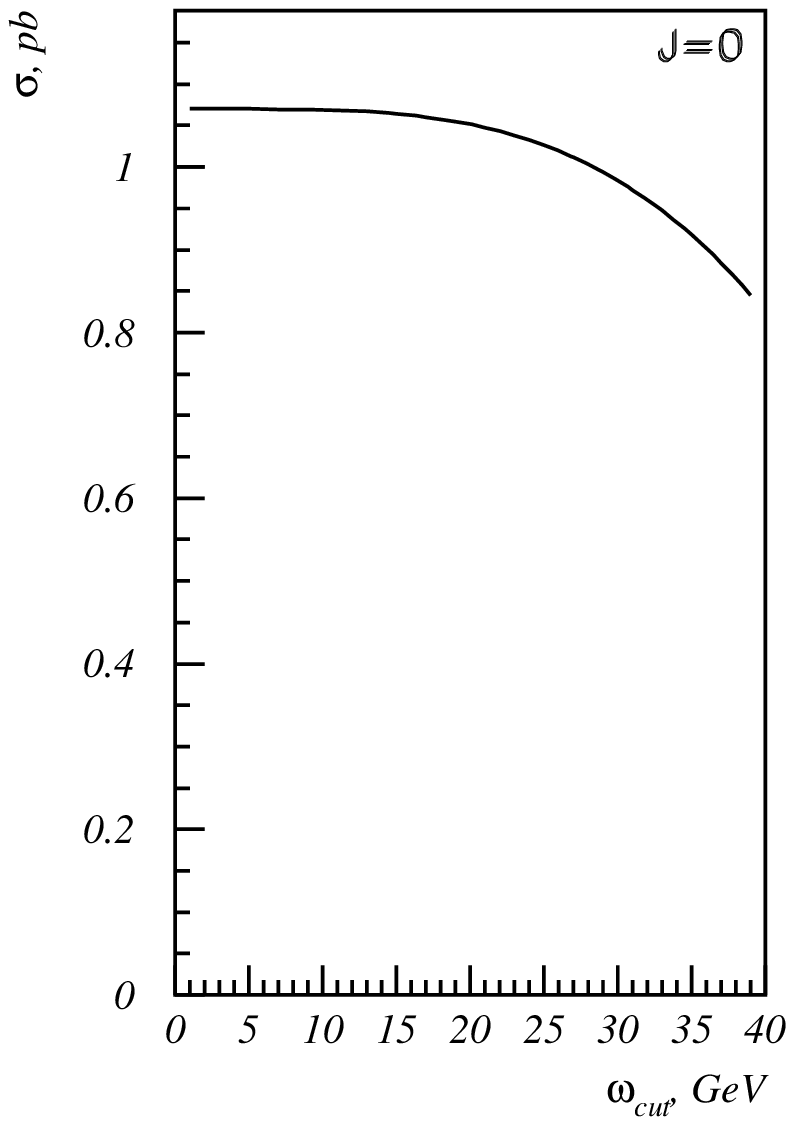}
\includegraphics[width=0.32\linewidth, bb=8 21 234 342, angle=0]{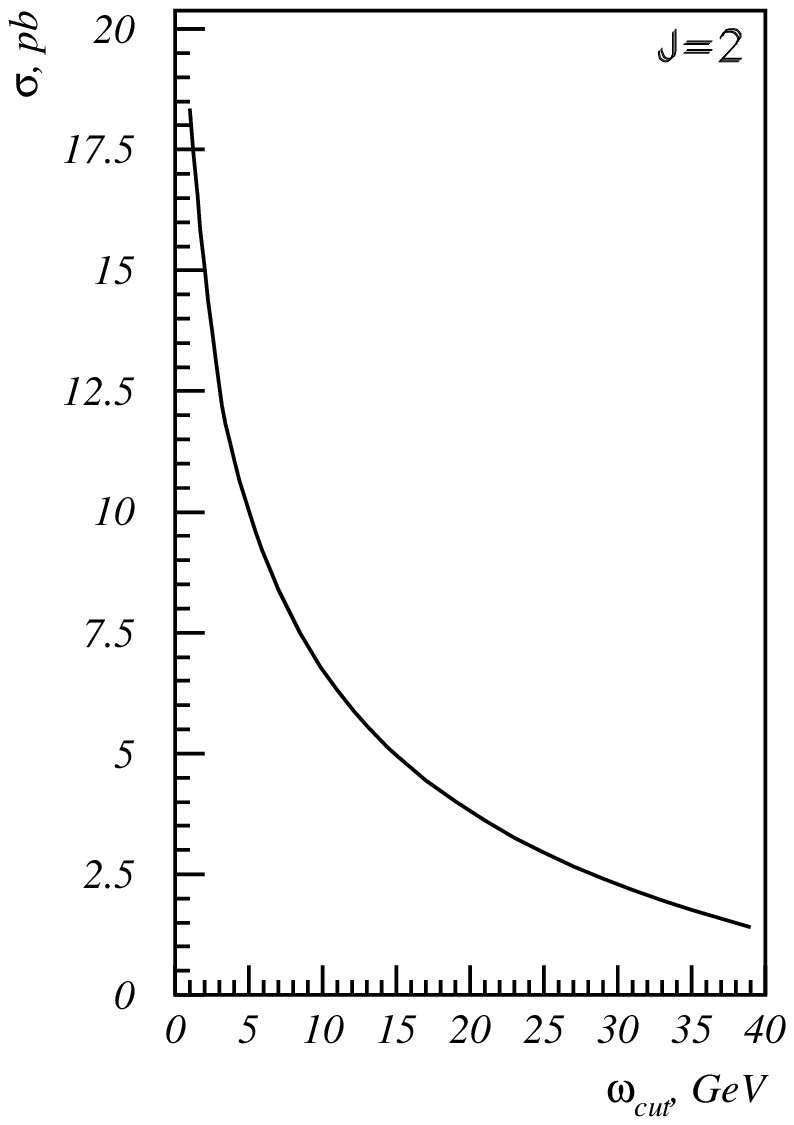}
\includegraphics[width=0.32\linewidth, bb=8 21 234 342, angle=0]{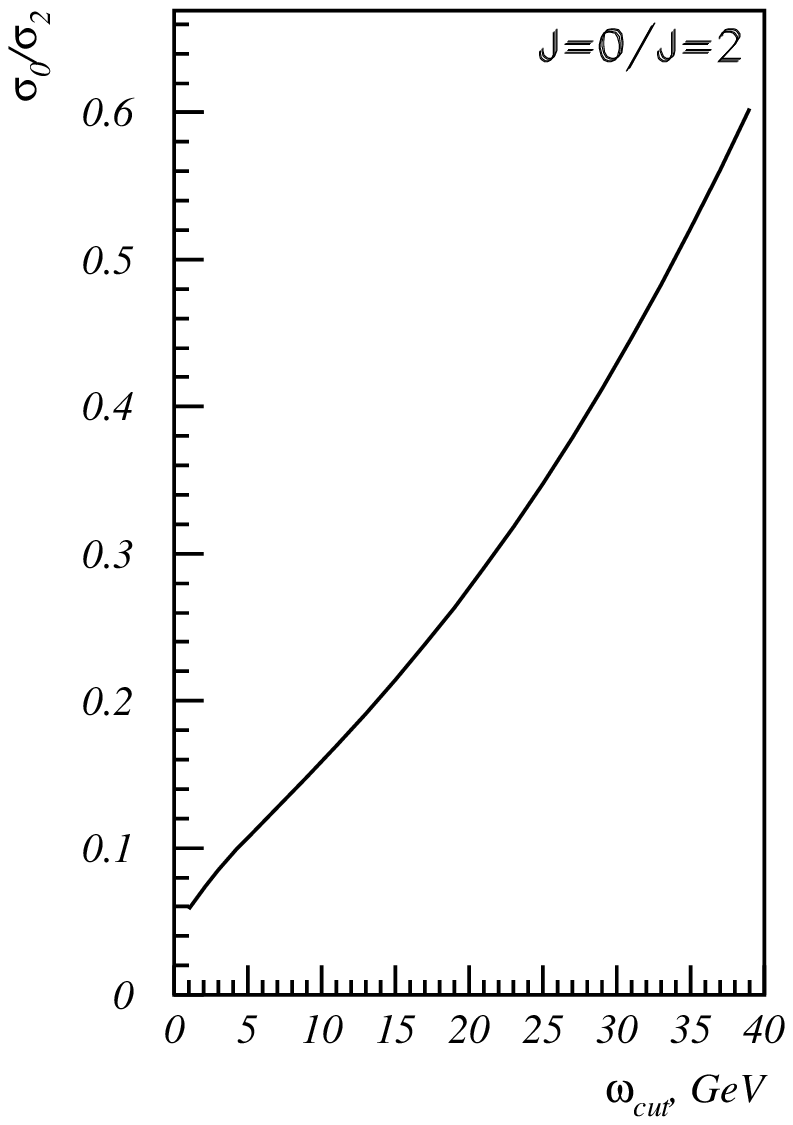}
\caption{
Cross sections for $\omega > \omega_{min}$ for $J\!=\!0$ and $J\!=\!2$ 
and their ratio.
}\label{fig_4}
\end{figure}

In fig. \ref{fig_4} we show the total cross section dependence on the 
$\omega$-cut and the ratio $\sigma_{J=0} \slash \sigma_{J=2}$.

A ratio $\sigma_{J=0} \slash \sigma_{J=2} > 0.5$
can be achieved without a large loss in $\sigma_{J=2}$.

\begin{figure}[p]
\leavevmode
\centering
\includegraphics[width=0.48\linewidth, bb=6 19 294 285, angle=0]{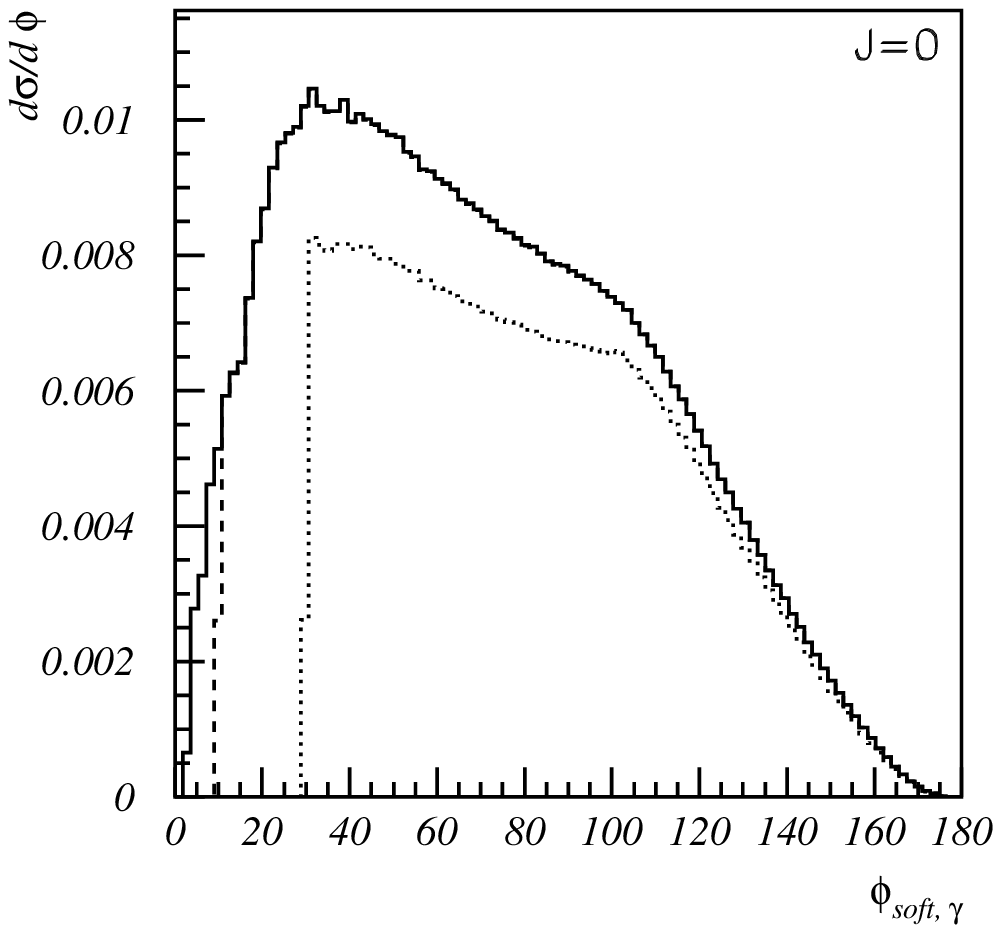}
\includegraphics[width=0.48\linewidth, bb=6 19 294 285, angle=0]{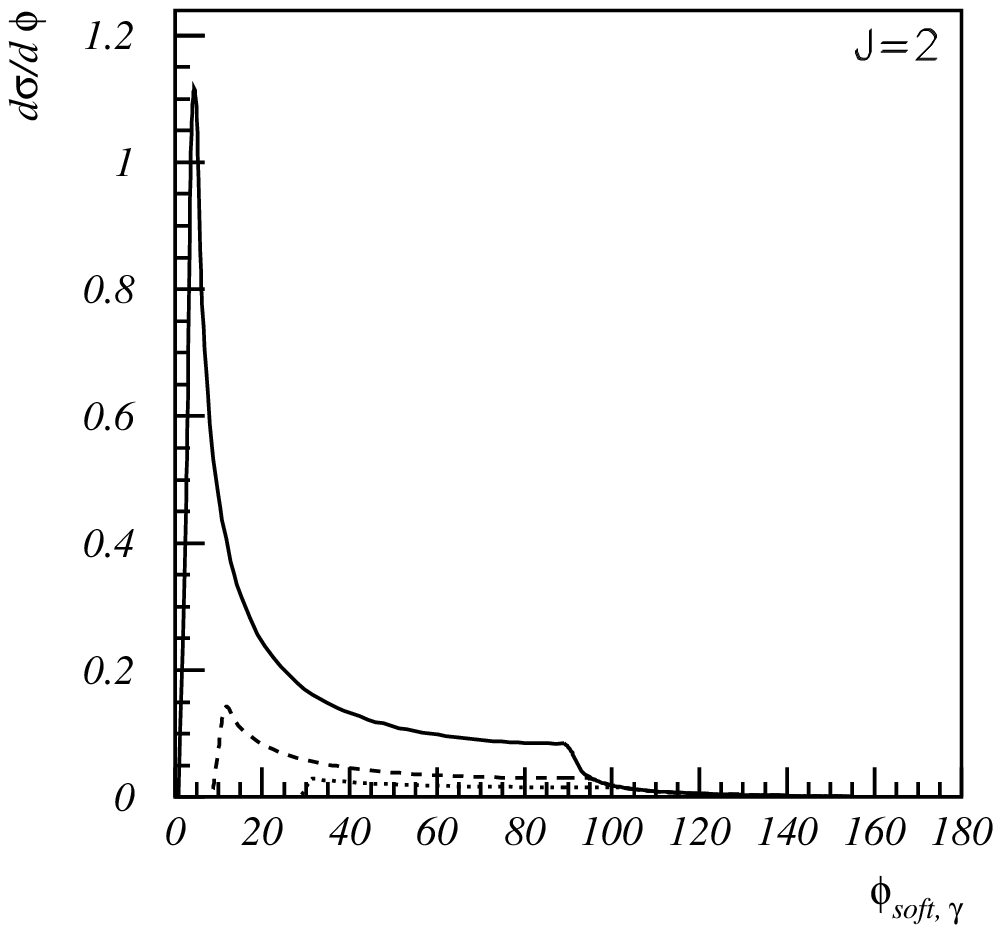}
\caption{
$d \sigma / d \varphi$ at $\sqrt{s} = 120 \GeV$ for $J=0$ and $J=2$
and various cuts:
$\Theta_{min} = 7^\circ$, $\varphi_{min} = 3^\circ$, $E_{\ell, min}=1 \GeV$,
$\omega_{min} = 1 \GeV$ (solid line);
$\Theta_{min} = 7^\circ$, $\varphi_{min} = 10^\circ$, $E_{\ell, min}=1 \GeV$, 
$\omega_{min} = 10 \GeV$ (dashed line);
$\Theta_{min} = 7^\circ$, $\varphi_{min} = 30^\circ$, $E_{\ell, min}=5 \GeV$, 
$\omega_{min} = 20 \GeV$ (dotted line).
}\label{fig_5}
\end{figure}

\begin{figure}[p]
\centering
\leavevmode
\includegraphics[width=0.32\linewidth, bb=8 21 228 343, angle=0]{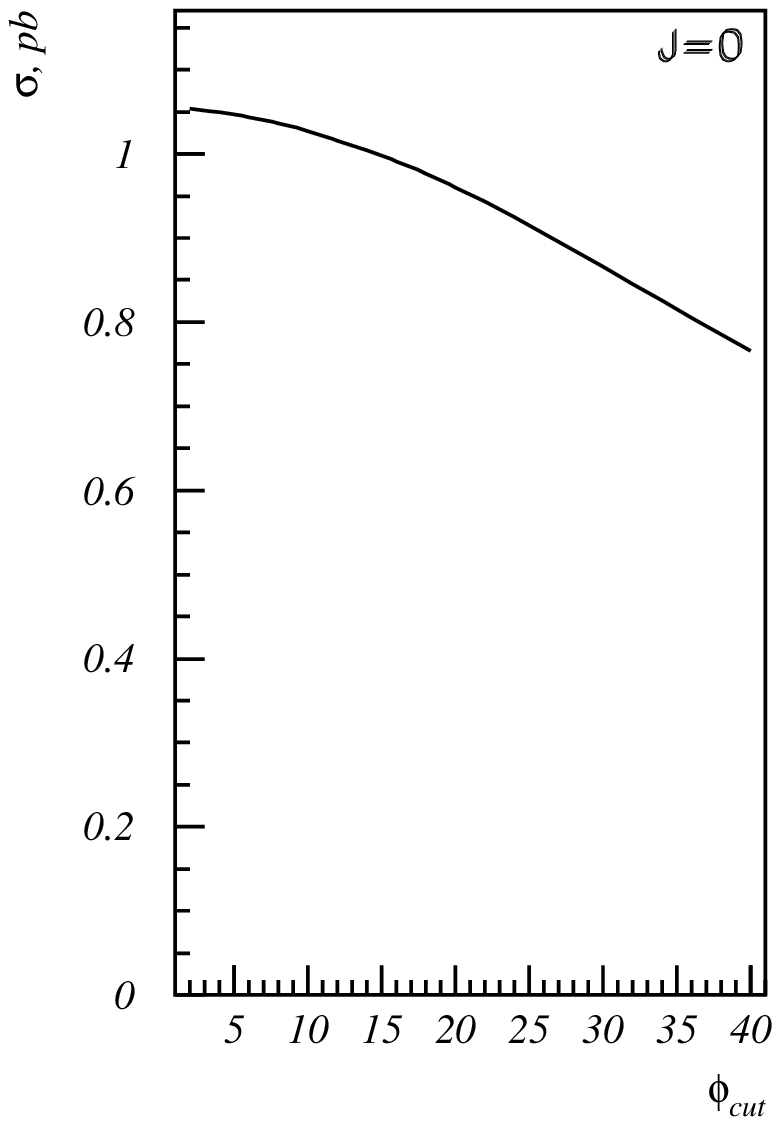}
\includegraphics[width=0.32\linewidth, bb=8 21 228 343, angle=0]{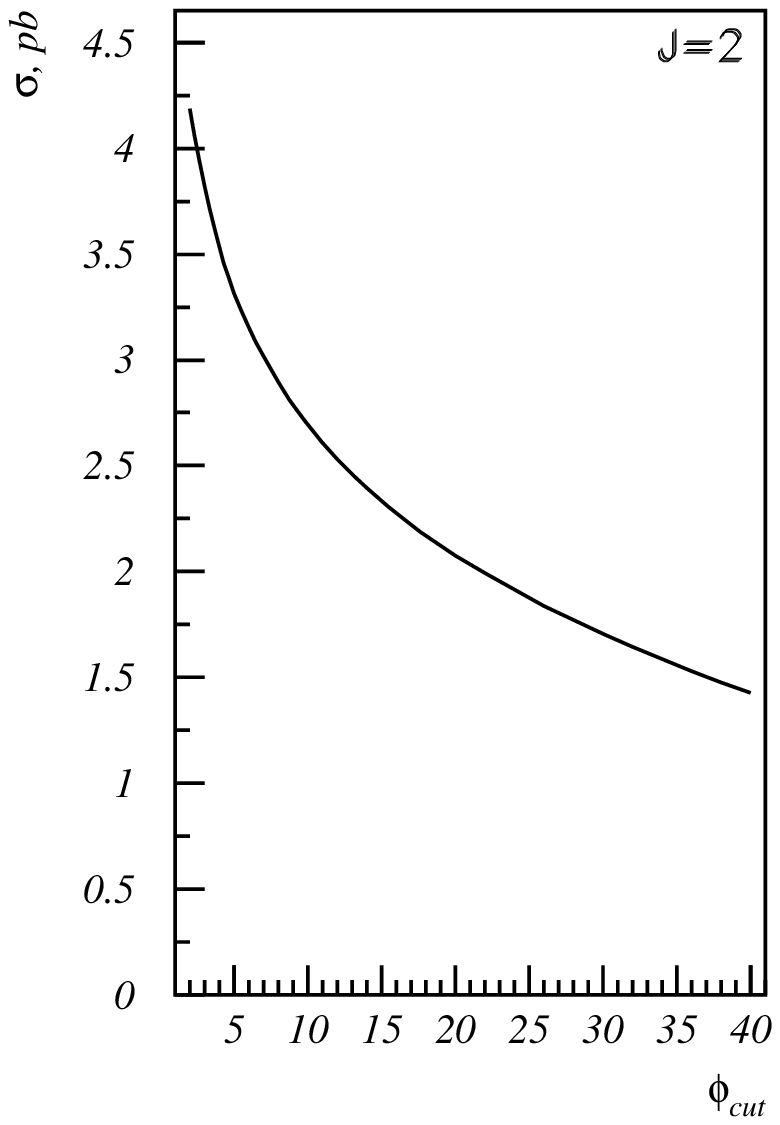}
\includegraphics[width=0.32\linewidth, bb=8 21 228 343, angle=0]{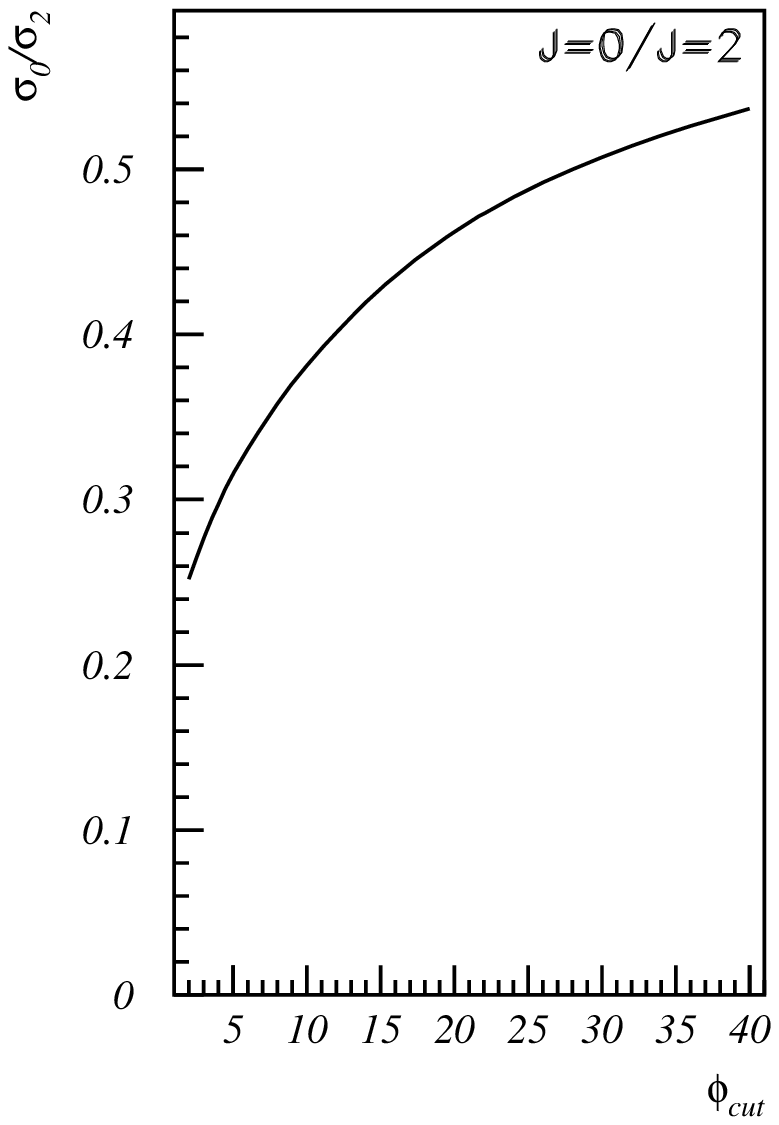}
\caption{
Cross sections for $\varphi > \varphi_{min}$ for $J\!=\!0$ and $J\!=\!2$ 
and their ratio.
}\label{fig_7}
\end{figure}

In figure \ref{fig_5} the minimum angle between the photon and one of the 
two leptons, $\varphi$, is shown for both beam polarisations.
As already shown for the photon energy for $J=2$ one can see the typical final 
state radiation pattern with the colinear and infrared divergencies while 
for $J=0$ large angles are preferred. The total cross sections with
$\varphi > \varphi_{min}$ for $J=0$ and $J=2$ and their ratio are 
shown in figure \ref{fig_7}.


\begin{figure}[p]
\centering
\leavevmode
\includegraphics[width=0.32\linewidth, bb=8 21 229 342, angle=0]{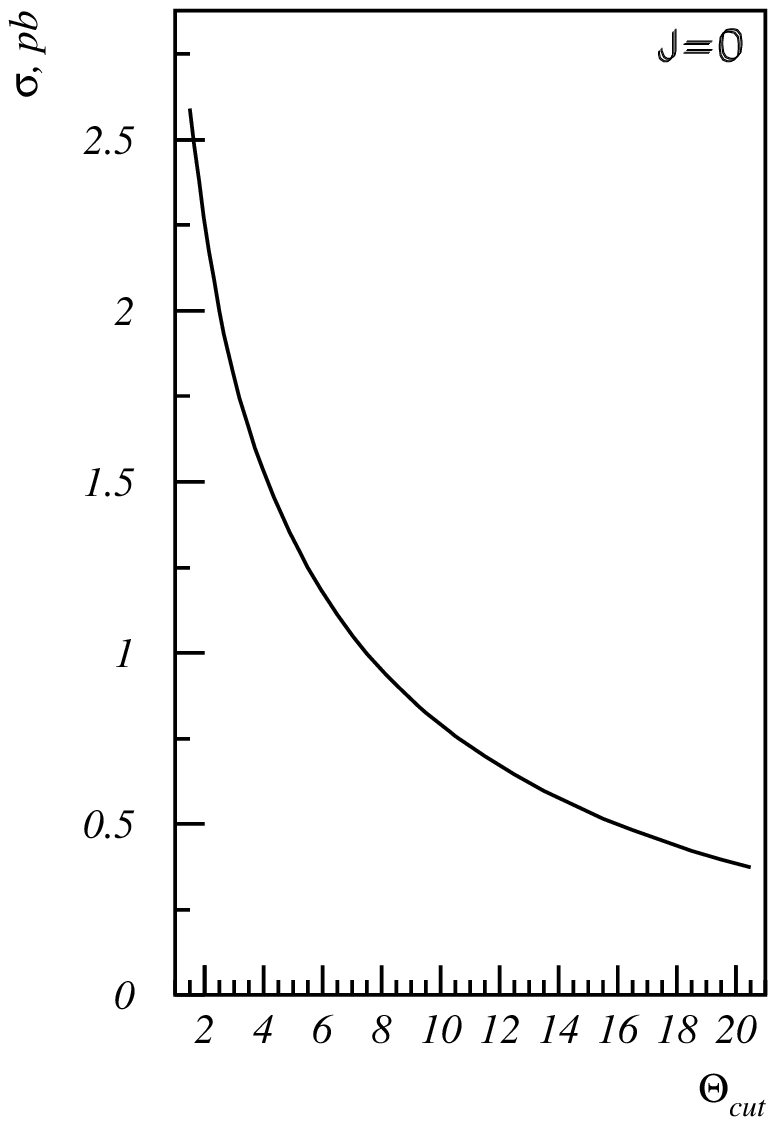}
\includegraphics[width=0.32\linewidth, bb=8 21 229 342, angle=0]{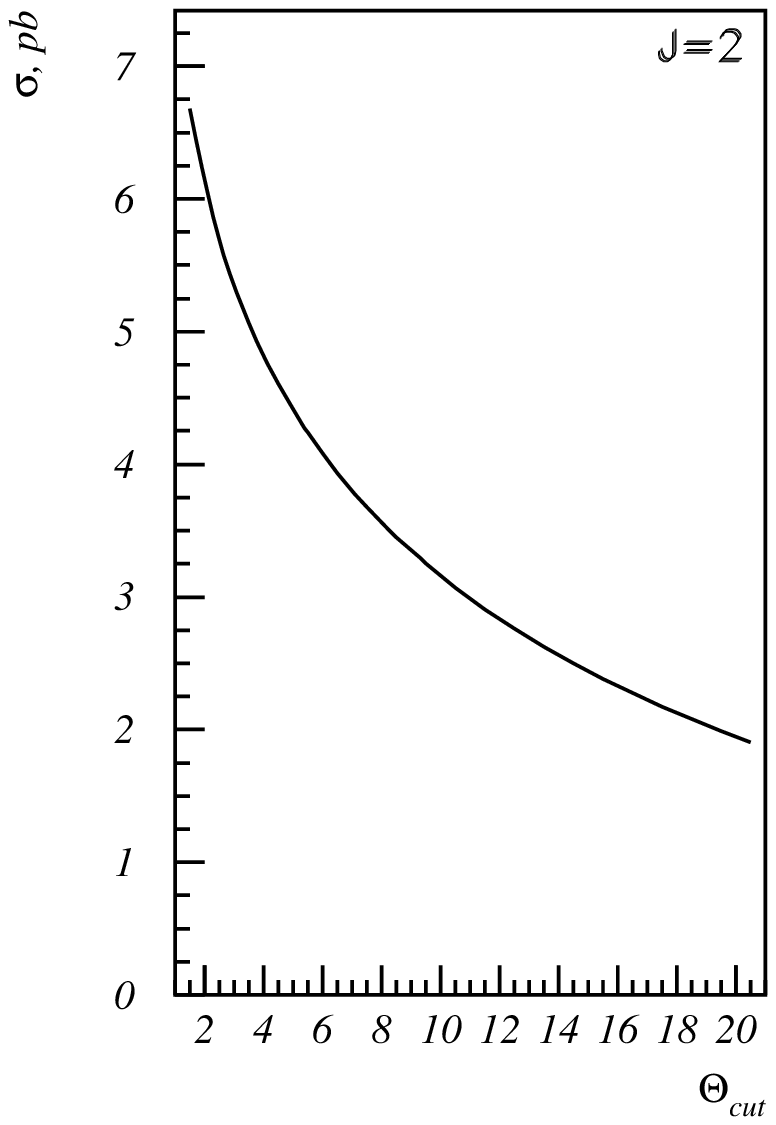}
\includegraphics[width=0.32\linewidth, bb=8 21 229 342, angle=0]{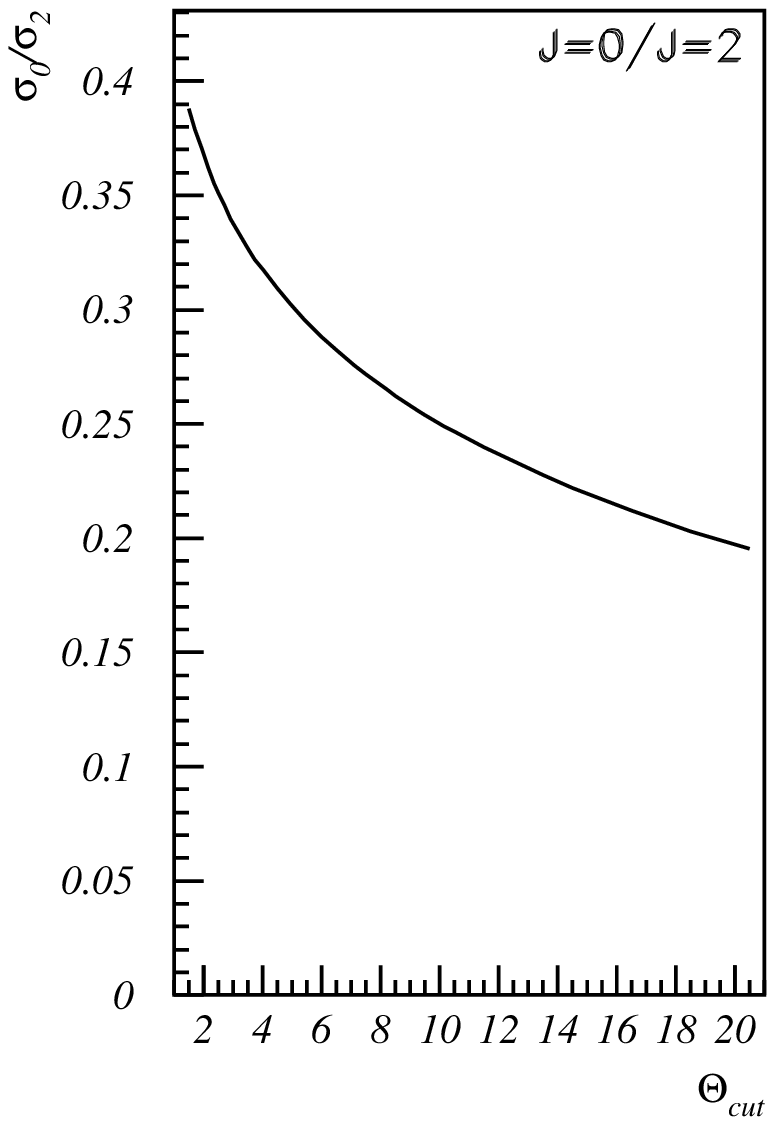}
\caption{
Cross sections for $\Theta > \Theta_{min}$ for $J\!=\!0$ and $J\!=\!2$ 
and their ratio.
}\label{fig_6}
\end{figure}

Figure \ref{fig_6} shows
the two total cross sections and their ratio as a function of
$\Theta_{min}$.
It is interesting to note that a low cut on $\Theta$ is not only needed to
get a large cross section but also enhances the $J=0$ component.

\begin{figure}[p]
\leavevmode
\centering
\includegraphics[width=0.48\linewidth, bb=8 23 287 285, angle=0]{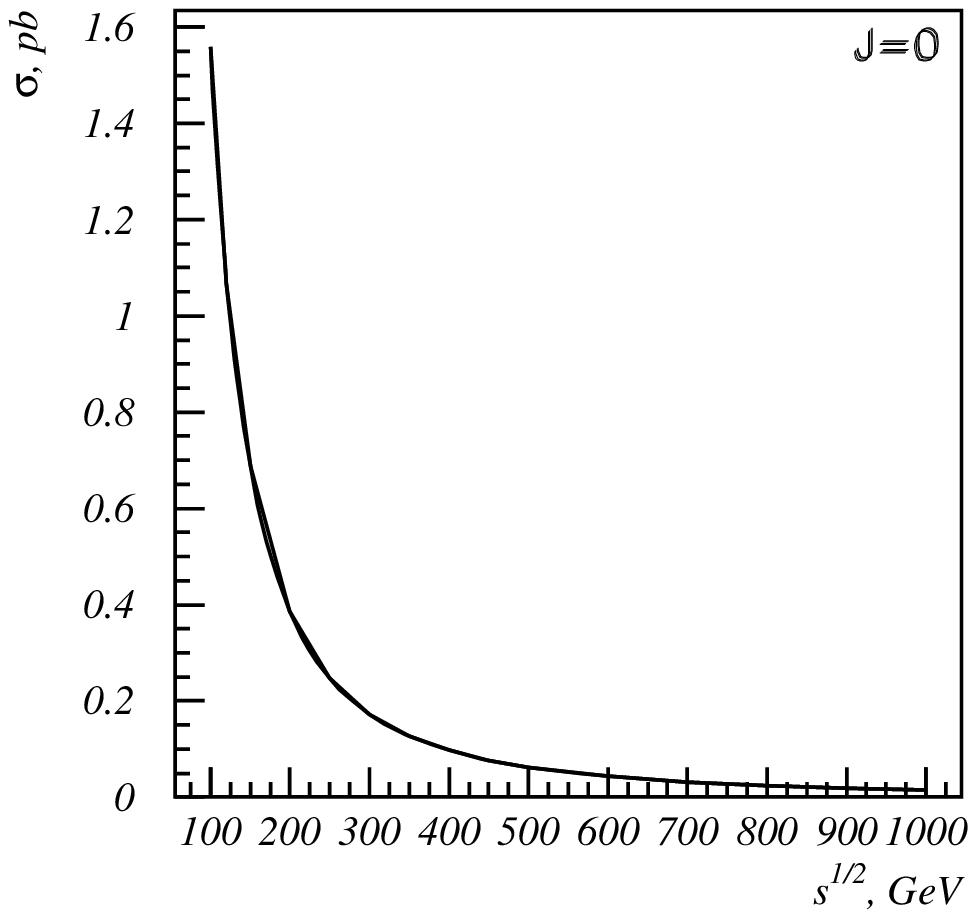}
\includegraphics[width=0.48\linewidth, bb=8 23 287 285, angle=0]{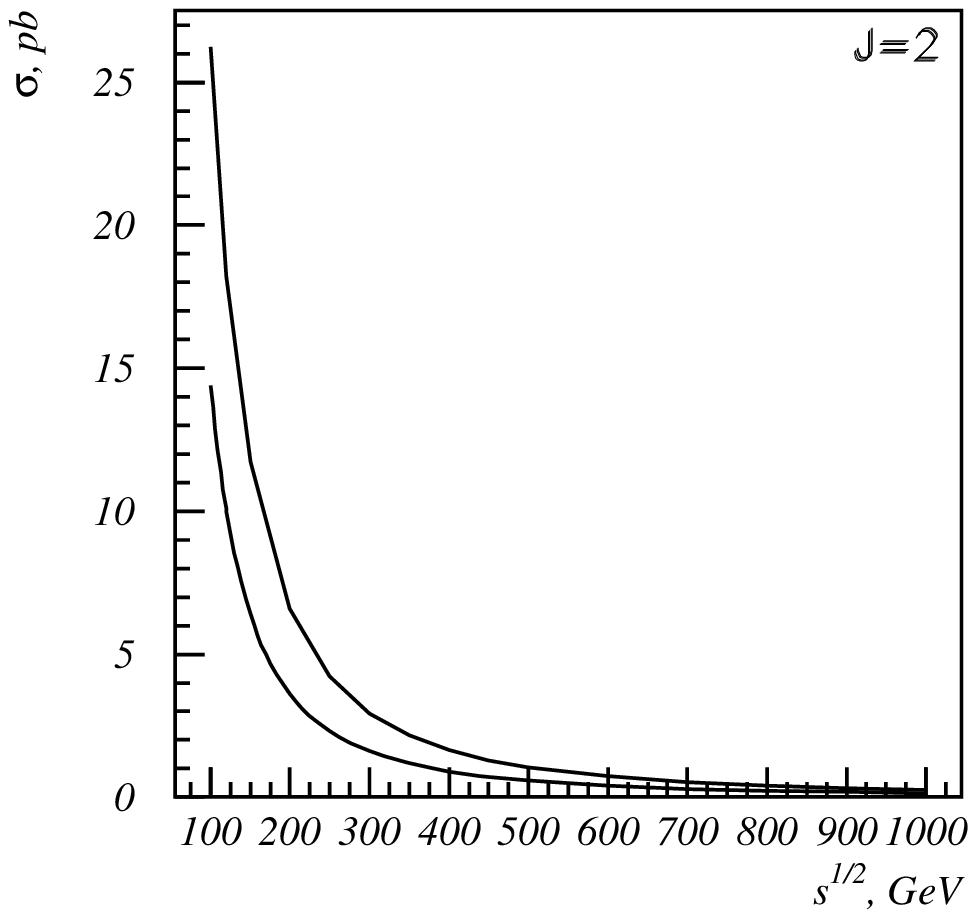}
\caption{
The dependence of the total cross section on $\sqrt{s}$.
The cuts are:
$\Theta_{min} = 7^\circ $, $\varphi_{min} = 3^\circ $,
$E_{\ell, min} = 1 \GeV$, $\omega_{min} = 1 \GeV\cdot (\sqrt{s}/120 \GeV )$ 
(upper line) and $\omega_{min} = 5 \GeV \cdot (\sqrt{s}/120 \GeV )$ 
(lower line).
}\label{fig_8}
\end{figure}

The total cross section dependence
on the centre of mass energy is shown in figure \ref{fig_8}.
It follows very well a $1 \slash s$ distribution.

All results presented in this section agree with an independent
analysis using Whizard \cite{whizard}. For $\sqrt{s} = 400 \GeV$ the results 
agree also with a calculation using CompHEP \cite{telnov}.

\section{Luminosity measurement}

The final cuts have been chosen to be
\begin{eqnarray*}
\Theta_{min} & = & 7^\circ,\\
\omega_{min} & = & 20 \GeV,\\
E_{\ell,min} & = & 5 \GeV,\\
\varphi_{min} & = & 30^\circ.
\end{eqnarray*}
These cuts simultaneously enhance the $J=0$ cross section compared to
the $J=2$ one and ensure that the events can be identified cleanly with
the detector.
With these cuts the final cross sections are
\begin{eqnarray*}
\sigma (J=0) & = & 0.82 \pb\\
\sigma (J=2) & = & 1.89 \pb
\end{eqnarray*}

For the standard beam and laser parameters the fraction of the $J=0$
polarisation in the total
luminosity is of the order $95\%$, so that the selected 
$\gamma \gamma \rightarrow \ell^+ \ell^- \gamma$ sample has a purity
of around 90\%. The $J=2$ background can be measured accurately with 
$\gamma \gamma \rightarrow \ell^+ \ell^-$ events.
For an electron beam energy of $E_e = 100 \GeV$ the TESLA design
luminosity for the high energy part of the beam is
${\cal L} (\sqrt{s'} > 0.8 \sqrt{s'_{\rm max}})  =  4.8 \cdot 10^{33} \lunit$
and for a window of $\pm 2 \GeV$ around a Higgs mass of $120 \GeV$ it
is ${\cal L} (\MH \pm 2 \GeV)  =  7 \cdot 10^{32} \lunit$
\cite{ggtdr,cain}.
In a two year run ($2 \cdot 10^7 {\rm s}$) using muons only this
leads to a statistical precision on the luminosity measurement of
\begin{eqnarray*}
\frac{\Delta {\cal L}}{{\cal L}} 
\left(\sqrt{s'} > 0.8 \sqrt{s'_{\rm max}}\right) 
& = & 0.4 \% \\
\frac{\Delta {\cal L}}{{\cal L}}
\left(\MH \pm 2 \GeV\right) 
& = & 1.0 \% .
\end{eqnarray*}
Adding electrons these precisions improve by a factor $1/\sqrt{2}$.

The size of the mass window around $\MH$ that can be used for
luminosity determination depends on the confidence one has in the
luminosity spectrum once the data are available. Studies with 
Circe2 \cite{circe} and Cain \cite{cain} indicate that the
differential luminosity in a $2 \GeV$ window around the maximum
changes by less than 1\% so that this window is certainly safe.
Studies with the fast simulation program Simdet \cite{simdet} show
that the invariant mass resolution of the detector for the accepted
events is around $1 \GeV$, consistent with a $2 \GeV$ mass window.

\section{Conclusions}

The differential luminosity of a photon collider running with $J=0$ at
a $\gamma \gamma$-centre of mass energy around $120 \GeV$ to produce
light Higgses can be measured with an accuracy around 0.7\% in a two
years run. The uncertainty of the event rate 
$\gamma \gamma \rightarrow {\rm H} \rightarrow \bb$ in the same
running time will be around 1.5\% \cite{piotr,aura} and the one of the
branching ratio ${\rm BR ( H} \rightarrow \bb)$ from $\ee$ running
will be around 2.5\% \cite{phystdr}, so that the uncertainty on the
partial width $\Gamma( {\rm H} \rightarrow \gamma \gamma)$ will not be
limited by the error on the luminosity.

\end{document}